# Quantifying Algorithmic Biases over Time

**Vivek K. Singh**
School of Communication and Information
Rutgers University, NJ
v.singh@rutgers.edu

**Ishaan Singh**
Department of Computer Science
Rutgers University, NJ
is393@scarletmail.rutgers.edu

**ABSTRACT**

Algorithms now permeate multiple aspects of human lives and multiple recent results have reported that these algorithms may have biases pertaining to gender, race, and other demographic characteristics. The metrics used to quantify such biases have still focused on a static notion of algorithms. However, *algorithms evolve over time*. For instance, Tay (a conversational bot launched by Microsoft) was arguably not biased at its launch but quickly became biased, sexist, and racist over time. We suggest a set of intuitive metrics to study the variations in biases over time and present the results for a case study for genders represented in images resulting from Twitter image search for #Nurse and #Doctor over a period of 21 days. Results indicate that biases vary significantly over time and the direction of bias could appear to be different on different days. Hence, one-shot measurements may not suffice for understanding algorithmic bias, thus motivating further work on studying biases in algorithms over time.

**INTRODUCTION**

Understanding algorithms in search engines and social media sites has been a topic of major interest in social computing research. There is now consensus that platforms cannot be thought of as neutral and that they play an important role in shaping the perceptions of the individuals using the platforms [1, 3, 5, 9]. In fact, some efforts in critical algorithm studies have described algorithms as "the new power brokers in society" [7, 8]. Besides critical analysis, multiple researchers are now analyzing the reasons behind bias and proposing methods to counter bias [8, 6, 9]. While there have been several recent empirical studies on quantifying algorithmic biases there has been an (implicit) assumption that the biases will be measured once or a small number of times. Hence, while multiple metrics such as disparate impact, equality of opportunity etc. have been proposed to measure the differences between privileged and unprivileged communities (e.g. male and female persons [1, 6, 10]) there have been no attempts that explicitly assume that the algorithms will be continuously evolving and try to quantify the temporal dynamics of the biases.

This is an important gap as **real-world algorithms evolve continuously.** Reinforcement Learning – a major branch of machine learning – is designed on the idea of continuously learning from newer data [2] and we have seen multiple public facing algorithms evolve continuously. Besides, Tay (a bot launched by Microsoft), which famously became more biased over time [3], Google autocomplete is known to learn from user input continuously [4] and YouTube recommendations have been reported to become more extreme over time [5]. Further, differentiating between the bias in data and bias in algorithm is likely futile because of the complex interplay between data and algorithm both feeding off each other [3]. To move the conversation forward, here we present a few metrics that we found to be useful for quantifying temporal variations in bias over time. We also illustrate the use of these metrics based on a case study for genders represented in results for Twitter image search for #Doctor and #Nurse over a period of 21 days. Results indicate that the levels of bias vary significantly over time (even more than the average reported bias) and hence suggest that one-time measurements of bias may be incomplete and even unreliable.

**CASE STUDY**

Here, we study the variations in bias across gender in Twitter image search over 21 days. Data were collected for a period of 3 weeks from June 2[nd] to June 22[nd,] 2019 daily at 6 p.m. in incognito mode. Taking inspiration from Kay et al.'s work [1], we define bias as the percentage difference in the representation of male and female persons in the image search results for "#Doctor" and "#Nurse." (Note that we consider the use of binary gender to be a limitation of this work.) One of the co-authors manually labeled the first 200 images for each day for each search term under different categories. If the protagonist (doctor, nurse) was easily identifiable we labeled their gender as 'Male', 'Female,' or 'Can't Say (Gender).' If the protagonist was not easily identifiable then that image was labeled 'Can't Say'. In 'Not Human' category we put all the images which don't have any humans in it. Hence, the dataset included labels for 'Male',' Female', 'Can't Say (Gender)', 'Can't identify protagonist', and 'Not Human'. For images which had multiple persons of different genders, the score for that image was split between genders in the ratio of presence. For example, if in an image we have 3 humans and 2 of them are male and 1 of them is female then we allocate 0.66 score to male category and 0.33 to Female category. See Table 1 for data for #Doctor search results over 21 days.

Table 1: #Doctor Data Set

| #Doctor  | Male  | Female | Can't identify protagonist | Not Human | Can't Say Gender |
|----------|-------|--------|----------------------------|-----------|------------------|
| 06-02-19 | 18.83 | 10.17  | 79                         | 88        | 4                |
| 06-03-19 | 19    | 20     | 72                         | 86        | 3                |
| 06-04-19 | 22.81 | 16.98  | 60                         | 97        | 4                |
| 06-05-19 | 18    | 19     | 67                         | 93        | 3                |
| 06-06-19 | 16.8  | 15.2   | 77                         | 88        | 3                |
| 06-07-19 | 15.6  | 19.4   | 72                         | 90        | 3                |
| 06-08-19 | 20.6  | 17.4   | 67                         | 92        | 3                |
| 06-09-19 | 18.1  | 19.9   | 68                         | 89        | 5                |
| 06-10-19 | 23    | 22     | 54                         | 96        | 5                |
| 06-11-19 | 17.3  | 23.7   | 64                         | 94        | 1                |
| 06-12-19 | 16.66 | 18.34  | 78                         | 84        | 3                |
| 06-13-19 | 19.6  | 19.4   | 81                         | 76        | 4                |
| 06-14-19 | 14.5  | 13.5   | 120                        | 49        | 3                |
| 06-15-19 | 27.5  | 23.5   | 70                         | 74        | 5                |
| 06-16-19 | 26.84 | 18.16  | 69                         | 81        | 5                |
| 06-17-19 | 26.32 | 25.68  | 73                         | 71        | 5                |
| 06-18-19 | 27    | 20     | 78                         | 75        | 5                |
| 06-19-19 | 25.66 | 20.34  | 73                         | 79        | 2                |
| 06-20-19 | 27.7  | 17.3   | 72                         | 79        | 4                |
| 06-21-19 | 19.7  | 20.3   | 61                         | 93        | 6                |
| 06-22-19 | 19    | 14     | 63                         | 100       | 4                |

Following the principle of Disparate Impact [6] and taking inspiration from Kay et al.'s [1] work we quantify bias as the difference in the percentage representation of men and women in image search results. Hence,

Bias = $Male\ Representation\% - Female\ Representation\%$.

Prior literature suggests that such differences in representations effect the perceptions of users regarding the professions and could reinforce stereotypes and create negative impressions [1]. Assuming equal representation, this number should be close to zero, but like many previous works, we found this number to be non-zero. Additionally, we found it to vary significantly over time.

**METRICS: QUANTIFYING DYNAMICS OF BIAS OVER TIME**
We consider bias scores to be a time series and for which it makes sense to quantify the overall rate of change, its significant inflection points, range, and the average tendency. Hence, we adopted the following metrics to quantify its evolution over time.

**(1) Rate of Change**

Rate of change (ROC) is the rate at which the bias is increasing or decreasing within observed time frame.

$$\text{ROC} = \frac{\text{Bias at last obsevation} - \text{Bias at 1st observation}}{\text{Total number of Observations}}$$

If we get the negative value that means that the algorithm is becoming less biased over time, else the opposite is happening.

**(2) Max and Min Bias Range**

Max Bias and Min Bias help us to determine the maximum and minimum level of bias observed. These provide the upper and lower bounds of bias that one could observe in the algorithm's results. These features could be combined to obtain the range of bias.

$$\text{Range} = \text{Max}(Bias) - \text{Min}(Bias)$$

This range helps to get a sense of the overall fluctuation and also interpret the validity of any specific observation regarding bias.

**3) Root Means Square of Bias**

Root mean square of bias (RMSB) provides an estimate of the average level of bias observed over the considered time period.

$$\text{RMSB} = \sqrt{\frac{bias\_1^2 + bias\_2^2 + \cdots + bias\_n^2}{n}}$$

Where *bias_i* is the bias measured in the i[th] observation and there are *n* observations in total.

**RESULTS**

The results of the image analysis for each day for the #Doctor search are summarized in Table 1. Similar data were obtained for #Nurse. To get the bias score, we first computed the relative representation of males and females in the search results as follows.

Male Representation % = $\frac{Male}{Male+Female} \times 100$ and Female Representation % = $\frac{Female}{Female+Male} \times 100$

Bias was computed as Bias = $Male\ Representation\% - Female\ Representation\%$. The resulting time series for bias in the case of #Doctor and #Nurse are shown in Figures 1 and 2 respectively. The computation results for the three temporal metrics of bias are shown in Tables 2-4.

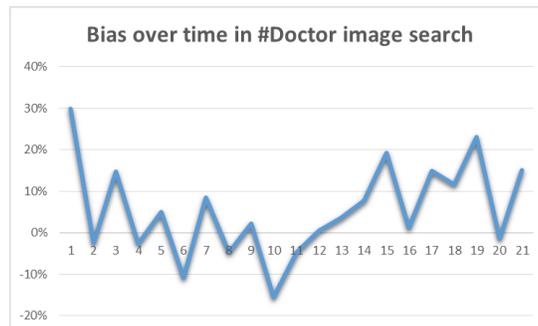

**Figure 1: Temporal variation in bias in #Doctor image search results on Twitter. Values indicated are the percentage differences in the relative representation of males and females in the results.**

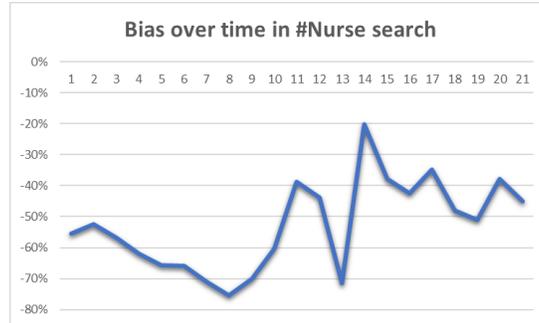

**Figure 2:** Temporal variation in bias in #Nurse image search results on Twitter. Values indicated are the percentage differences in the relative representation of males and females in the results.

**(1) Rate of Change**

| Categories | Delta (1ˢ day) | Delta (2nd Day) | Rate of change |
|---|---|---|---|
| Doctor | 30% | 15% | -0.71% |
| Nurse | -56% | -45% | -0.52% |

Table 2: Rate of change

As we can see in Table 2, the bias level i.e. the delta in the representation of men and women for #Doctor has reduced by 15% over the period of 21 days. This indicates that bias has been reducing over time at the rate of -0.71% per day. The same rate was found to be -0.52% per day for #Nurse.

**(2) Max & Min Bias Range**

Table 3: Max and Min Delta Range

| Categories | Max delta | Min Delta | Range |
|---|---|---|---|
| Doctor | 30% | -16% | 46% |
| Nurse | -20% | -75% | 55% |

As can be seen in Table 3, for #Doctor the results some time over represent males by as much as 30% and sometimes underrepresent males by as much as 16%. Hence, the direction of bias would appear to be different to users on different days. This motivates further validity analysis of multiple one-shot bias results studied in the past. This also has implications for the kind of counter steps to be undertaken as they are likely to differ quite significantly based on the direction of overrepresentation. The overall range for bias was 46% for #Doctor and 55% for #Nurse. Hence, the level of bias fluctuates quite significantly over time.

**(3) Root Mean Square of Bias**

Table 4: Root Mean Square of Bias

| Categories | Delta |
|---|---|
| Doctor | 12.34% |
| Nurse | 54.61% |

As we can see from results in Table 4, the Root Mean Square of Bias for #Doctor was 12.34% and that for #Nurse was 54.61%. In general, the results for #Nurse were more skewed towards one gender than the results for #Doctor. Note that this average bias level was lesser than the fluctuation in levels of bias (Range) as shown above, which again motivates detailed validity analysis of one-shot bias results henceforward.

## CONCLUSION

This work moves the conversation on algorithmic bias forward by arguing for ways to quantifying bias over time. The early results presented in this work suggest that *direction* of bias could vary over time and hence the deciding on interventions based on a single or a small number of measurements may not be appropriate. As algorithms evolve over time, it may be important to understand not only their average behavior but also the major inflection points and the rates of change. More empirical work including metrics, case studies, and conceptual frameworks are needed to study biases in algorithms that vary over time.